\documentclass{ere04}

\usepackage{graphicx}    

\usepackage{amsmath,amssymb}

\def\beq{\begin{equation}}
\def\eeq{\end{equation}}

\def\bea{\begin{eqnarray}}
\def\eea{\end{eqnarray}}
\def\ba{\begin{array}}                  
\def\ea{\end{array}}

\begin{document}

\title*{On non-singular inhomogeneous cosmological models}
\author{L. Fern\'andez-Jambrina\inst{1}\and
L.M. Gonz\'alez-Romero\inst{2}}
\institute{E.T.S.I. Navales, Universidad Polit\'ecnica de Madrid, Arco
de la Victoria s/n, E-28040 Madrid, Spain.
\texttt{lfernandez@etsin.upm.es} \and Departamento de F\'{\i}sica
Te\'orica II, Facultad de F\'{\i}sica, Avenida Complutense s/n,
E-28040 Madrid, Spain.  \texttt{mgromerod@fis.ucm.es}}

\maketitle

In this talk we would like to review recent results on non-singular
cosmological models.  It has been recently shown that among stiff
perfect fluid inhomogeneous spacetimes the absence of singularities
is more common than it was expected in the literature.  We would like
to generalize these results and apply them to other matter sources.

\section{Introduction}
\label{sec:1}

The interest for non-singular cosmologies has risen since it was shown
that among inhomogeneous stiff fluid spacetimes there was a regular
family depending on two nearly arbitrary functions \cite{wide}. 

Since the publication of the first known non-singular perfect fluid
cosmological model by Senovilla \cite{seno}, the number of regular
spacetimes had not increased a lot.  Senovilla and Ruiz enlarged the
family \cite{ruiz} two years afterwards.  All of them had the equation
of state of an incoherent radiation fluid.  The other regular models
that were found either had no equation of state \cite{Sep} or were for
stiff fluids \cite{dadhich,Diag,Jerry,leo}.

As it has been said, the role of the stiff fluid equation of state in
the absence of singularities is well known, since the gradient of
pressure prevents the formation of singularities. In fact, it is a
limiting case between causal and non-causal linear equations of state,
since the velocity of sound for the stiff fluid is equal to the velocity of 
light. On the contrary, dust models, with vanishing pressure, are all
singular.

Therefore, Senovilla's original solution is intriguing for the fact
that it contains incoherent radiation as a matter content, whereas
every other barotropic regular model has been found for a stiff
equation of state.  Furthermore, in the generalization of the model by
Senovilla and Ruiz incoherent radiation spacetimes are the only ones that are
non-singular and they trace the separatrix between models which have
Ricci and Weyl singularities.

In this talk we pursue to analyze the role of radiation in the absence
of singularities in inhomogeneous cosmological models.  To this aim we
shall consider pure radiation cosmological models in the framework of
orthogonally-transitive, mutually orthogonal Abelian $G_{2}$
spacetimes, where most regular perfect fluid models have been found so
far.  There are spherically symmetric non-singular models
\cite{grg}, but the matter content is an anisotropic fluid.

In the next section we shall derive the equations for pure radiation
inhomogeneous cosmological models and in the last section we shall
analyze the geodesic completeness of such models. The results will be 
summarized in a conclusions section.

\section{Pure radiation inhomogeneous models}
\label{sec:2}

We start by defining the matter content of the models. The
energy-momentum tensor for pure radiation,

\begin{equation}
    T^{\mu\nu}=\Phi k^{\mu}k^\nu,
    \label{null}
\end{equation}
is written in terms of a null vector field describing the 4-velocity
of the radiation field $k$ and a positive function $\Phi$. It is
tempting to identify $\Phi$ with the energy density of the radiation
field, but it must be noticed that it is not unambiguously defined,
since it may be altered by rescaling the vector field $k$.

In full generality, conservation laws for this energy-momentum,

\begin{equation}
    T^{\mu\nu}_{\ ;\nu}=\Phi_{,\nu}k^\nu k^{\mu}+\Phi a^{\mu}+\Phi
    \Theta k^{\mu}=0,
    \label{conserved}
\end{equation}
are written in terms of the expansion $\Theta=k^\nu_{;\nu}$ and the acceleration,
$a^\mu=k^\mu_{;\nu}k^{\nu}$ of the vector field $k$. 

If $k$ were timelike instead of null, we could split the equation in
two components, parallel and orthogonal to $k$. But in the null case
there is a single component if $\Phi$ is non-zero.

As we have already stated, we require that the spacetimes allow an
orthogonally-transitive, mutually orthogonal $G_{2}$ group of
isometries. The metric,

\begin{equation}
ds^2=e^{2K}(-dt^2+dx^2)+\rho^2e^{2U}dy^2+e^{-2U}dz^2,\label{metric}
\end{equation}
is written in coordinates $y$, $z$, adapted to the isometry generators,
which do not appear in the metric functions $\rho$, $K$, $U$, and two
isotropic coordinates $t$ and $x$.  The function $K$ is just a
conformal factor and $\rho$ is the surface element of the transitivity
surfaces.  It is much related to the absence of singularities, since
all regular models have a spacelike gradient of $\rho$.  The
coordinates have the usual ranges.

We impose that $k$ has just components on $t$ and $x$,

\begin{equation}
    k=e^{-K}(\partial_{t}+\partial_{x}),
\end{equation}
for an outgoing radiation flux. For ingoing radiation we have just to 
change the plus sign for a minus in the sum.

In these coordinates the conservation equation can be integrated, 

\begin{equation}
\left(\partial_{t}+\partial_{x}\right)\left\{\Phi e^{2K}\rho\right\}=0
\Rightarrow 
    \Phi e^{2K}\rho=h(t- x),
    \label{cons2}
\end{equation}
in terms of one function $h$ of the outgoing null coordinate $t-x$.

The remaining Einstein equations in these coordinates, 

 \begin{subequations}
 \begin{eqnarray}
 &&
 U_{tt}-U_{xx}+\frac{1}{\rho}(U_{t}\rho_{t}-U_{x}\rho_{x})=0,\label{U}
 \\
 && \rho_{tt}-\rho_{xx}=0,\label{rho}
 \\
 &&
 K_{t}\rho_{x}+K_{x}\rho_{t}=\rho_{tx}+U_{t}\rho_{x}+U_{x}\rho_{t}+2
 \rho U_{t}U_{x} - h,\label{Kt}
 \\
 &&
 K_{t}\rho_{t}+K_{x}\rho_{x}=\frac{\rho_{tt}+\rho_{xx}}{2}+
 U_{t}\rho_{t}+U_{x}\rho_{x}
 +\rho\left(U_{t}^2+U_{x}^2\right)+h,\label{Kx}
 \end{eqnarray}
 \end{subequations}
may be reduced to a plane-wave equation for $U$ and a wave equation
for $\rho$. The equations for $K$ are a quadrature. The integrability 
condition for these equations is a consequence of the other
equations and therefore they can be integrated at the end. 

If we take $\rho$ as a coordinate, the system is further simplified.
There are three possibilities:

\begin{itemize}
    \item  If $\rho$ is a null coordinate, $\rho=\rho(t- x)$,
    $U=U(t-x)$ is also null and the only equation is a wave equation,
\begin{equation}
    K_{tt}-K_{xx}=0 \Rightarrow K(u,v)=f(u)+g(v),
\end{equation}
which may integrated in terms of null coordinates, $u=t-x$, $v=t+ x$. 
Just $f$ is constrained by the quadrature,

\begin{equation}
    f'=\frac{\rho''}{2\rho'}+U'+\frac{\rho U'^2}{\rho'}+\frac{h}{2\rho'},
    \label{f}
\end{equation} where the prima denotes derivation with respect to $u$.

After eliminating redundant functions by defining $\tilde u=\int du\,
e^{2f}$, $\tilde v=\int dv\, e^{2g}$, the metric is written in a
simple form,

\begin{equation}\label{lightmetric}
    ds^2=-2du dv+F^2dy^2+G^2dz^2,
\end{equation}
in terms of two arbitrary functions, $F$ and $G$, of $u$. We have
dropped the tilde for simplicity.

The function $\Phi$ must be positive,
\begin{equation}\label{phiphi}
    \Phi=-\left(\frac{F''}{F}+\frac{G''}{G}\right)\ge 0.
\end{equation}

These models have a five-dimensional group of isometries acting on
null orbits.  The vector field $\partial_{v}$ is covariantly constant
and these spacetimes contain plane-fronted gravitational waves with
parallel rays.
 
\item If $\rho$ is a spacelike coordinate, we choose $\rho=x$,

\begin{subequations}
\begin{eqnarray}
&&
U_{tt}-U_{xx}-\frac{U_{x}}{x}=0,\label{U2}
\\
&&
K_{t}=U_{t}+2x U_{t}U_{x} - h,\label{Kt2}
\\
&&
K_{x}=U_{x}+x\left(U_{t}^2+U_{x}^2\right)+ h,\label{Kx2}
\end{eqnarray}
\end{subequations}
and the equations are the same as for vacuum spacetimes except for the
function $h$, which appears in the metric as a conformal factor $-
H(t- x)$ added to $K$ \cite{rao}.

Since according to equation \ref{cons2} the function $h(t- x)$ must be positive
for positive $x$ and negative for negative $x$ we have to interpret
$x$ as a radial coordinate ranging from zero to infinity, although the
axis is not flat.  Therefore $H_{t}=h$ must be positive and $H$ is a
growing function.

\item Finally, if $\rho$ is a timelike coordinate, we take $\rho=t$,

\begin{subequations}
\begin{eqnarray}
&&
U_{tt}-U_{xx}+\frac{U_{t}}{t}=0,\label{U3}
\\
&&
K_{x}=U_{x}+2t U_{t}U_{x} - h,\label{Kx3}
\\
&&
K_{t}=U_{t}+t\left(U_{t}^2+U_{x}^2\right)+ h,\label{Kt3}
\end{eqnarray}
\end{subequations}
and the results are quite similar to the spacelike case. We shall not 
develop this case since it is singular at $t=0$.

\end{itemize}

\section{Geodesic completeness}
\label{sec:3}

There are at least three different definitions for singularities
\cite{HE}.  One of them states that a spacetime is singular if any of
the polynomial curvature scalars blow up at an event.  This definition
has the drawback that singular events may be hidden by conveniently
removing an open set aroung them. 

Therefore a more reasonable definition would be considering causal
observers on trajectories on the spacetime. For instance, restricting 
to observers in free fall, which move along timelike and null
geodesics, a singularity is encountered if their proper time cannot be
extended from minus infinity to infinity. This would mean that the
observer has left the spacetime in a finite proper time. Spacetimes
which do not suffer this inconvenience are called causally
geodesically complete. We shall adopt this definition for regular
spacetimes.

There is another definition of singularity which refers to general
causal observers, not necessary in free fall and therefore not
following causal geodesics. 

In order no analyze the regularity of our spacetimes we are lead to
study their causal geodesics. 

A geodesic of 4-velocity $u$,

\begin{equation}
    u^\mu=\dot x^\mu=\frac{dx^\mu(\tau)}{d\tau},
    \end{equation}
where $\tau$ is the proper time, satisfies the geodesic equations
written in the form,

\begin{equation}
    a^\mu=\ddot x^\mu+\Gamma^\mu_{\nu\rho}\dot x^\nu \dot x^\rho=0,
    \qquad
    -\infty<\tau<\infty,
\end{equation}
in terms of the Christoffel symbols for the metric of the spacetime.

Since these models have isometries, the geodesic equations become
simpler, because there are conserved quantities of geodesic motion
assigned to them. If $\xi$ is a generator of an isometry, the
quantity $X$,

\begin{equation}
    X:=g_{\mu\nu}\xi^\mu u^\nu,
    \end{equation}
is conserved along geodesics of 4-velocity $u$, 

\begin{equation}
D_{u}X=X_{,\sigma}u^{\sigma}=\left(g_{\mu\nu;\sigma}\xi^\mu u^\nu+g_{\mu\nu}\xi^\mu_{;\sigma}
u^\nu+g_{\mu\nu}\xi^\mu u^\nu_{;\sigma}\right)u^{\sigma}=0,
\end{equation}
since the metric $g$ is covariantly constant, the acceleration
$a^\nu=u^\nu_{;\sigma}u^{\sigma}$ of the geodesics is zero and the
generator satisfies the Killing equation,

\begin{equation}
    \xi_{\mu;\nu}+ \xi_{\nu;\mu}=0.
    \end{equation}
    
Besides, the character of the geodesic is preserved,

\begin{eqnarray}
    \delta:=-g_{\mu\nu}u^\nu u^\rho,
\end{eqnarray}
and equal to zero for null geodesics and to one for timelike geodesics.
In fact, geodesic equations merely state this fact.

All these conserved quantities will be useful.

We have classified the spacetimes by the character of the coordinate
function $\rho$:

\begin{itemize}
    \item The function $\rho$ is a null coordinate: We have seen that
    the vector fields $\{\partial_{v}, \partial_{y}, \partial_{z}\}$
    are generators of isometries and therefore provide constants of
    geodesic motion,

    \begin{eqnarray}
    V & = & \langle \partial_{v},u\rangle=
    g_{v\mu}u^\mu =-\dot u,\\
	Y & = & \langle \partial_{y},u\rangle=
	g_{y\mu}u^\mu= F^2\dot y,  \\
    Z & = & \langle \partial_{z},u\rangle=
    g_{z\mu}u^\mu= G^2\dot z.
    \end{eqnarray}

All these equations are integrable provided that $F^{-2}$ and $G^{-2}$ 
are integrable functions. We are left just with one equation for $\dot
v$, which we may derive directly from the expression for $\delta$,
without using the Christoffel symbols,

    \begin{equation}
    \frac{dv}{du} = \frac{1}{2V^2}\left\{\delta+Y^2F^{-2}+Z^2G^{-2}\right\}.
    \end{equation}

    These results are comprised in a simple theorem:

     \begin{theorem}
	 A spacetime endowed with a metric (\ref{lightmetric}) and
	 with a pure radiation field as matter content is causally
	 geodesically complete if and only if the metric components
	 $g^{yy}$ and $g^{zz}$ are integrable functions of the
	 null coordinate $u$.
    \end{theorem}

However, as it is explained in \cite{mpla}, the positiveness
requirement for the function $\Phi$ implies that the metric functions
are continuous but not differentiable.  

    \item  The function $\rho$ is a spacelike coordinate: In this case
    there are in general no additional isometries. However, there are 
    theorems specifically derived for this case \cite{manolo};

    \begin{theorem}
     A diagonal Abelian orthogonally transitive spacetime
    with spacelike orbits endowed with a metric in the form (\ref{metric}) with $C^2$
    metric functions  $K,U,\rho$, where $\rho$ has a spacelike gradient, is future causally geodesically complete
    provided that along causal geodesics:
    \begin{enumerate}
    \item For large values of $t$ and increasing $x$,
    \begin{enumerate}
    \item $(K-U-\ln\rho)_{t}+(K-U-\ln\rho)_{x}\ge 0$, and either
    $(K-U-\ln\rho)_{x}\ge 0$  or $|(K-U-\ln\rho)_{x}|\lesssim
    (K-U-\ln\rho)_{t}+(K-U-\ln\rho)_{x}$.
    \item $K_{t}+K_{x}\ge 0$, and either $K_{x}\ge 0$ or
    $|K_{x}|\lesssim K_{t}+K_{x}$.
    \item $(K+U)_{t}+(K+U)_{x}\ge 0$, and either $(K+U)_{x}\ge 0$
    or $|(K+U)_{x}|\lesssim (K+U)_{t}+(K+U)_{x}$.
    \end{enumerate}

    \item \label{tt} For large values of  $t$, a constant $b$ exists such that
    \\
    $\left.\begin{array}{c}K(t,x)-U(t,x)\\2\,K(t,x)\\K(t,x)+U(t,x)+\ln\rho(t,x)
    \end{array}\right\}\ge-\ln|t|+b.$

    \end{enumerate}

    For geodesics pointing to the past the theorem imposes the same
    conditions but reversing the sign in time derivatives.\end{theorem}

This result imposes certain restrictions on the metric functions:

\begin{theorem}
    A diagonal Abelian orthogonally-transitive spacetime with
    spacelike orbits with a metric in the form (\ref{metric}) and a
    spacelike $\rho$ and pure outgoing radiation as matter content is
    causally geodesically complete if:

    \begin{enumerate}
	\item  $x^{1-\varepsilon}|U_{x}\pm U_{t}| \not\to 0$ for large values
    of $|t|$ and $x$.

	\item  $H(t) \le \ln|t|+b$ and $U(t,0) \ge  b-\frac{1}{2}\ln|t|+
	H(t)$ for large values of $t$.

    \item $U(t,0) \ge  b-\frac{1}{2}\ln|t|+ H(t)/2$ for small values of $t$.
    \end{enumerate}
\end{theorem}

This result has obvious consequences.  The amount of energy that can
be radiated, which may be measured by $H$, must be bounded in order to
prevent the formation of singularities.  For ingoing radiation the
result is very similar.

\begin{theorem}
    A diagonal Abelian orthogonally transitive spacetime with
    spacelike orbits with a metric in the form (\ref{metric}) and a
    spacelike $\rho$ and pure ingoing radiation as matter content is
    causally geodesically complete if:

    \begin{enumerate}
    \item  $x^{1-\varepsilon}|U_{x}\pm U_{t}| \not\to 0$ for large values
    of $|t|$ and $x$.

    \item  $U(t,0)  \ge  b-\frac{1}{2}\ln|t|-H(t)/2$ for large values of $t$.

    \item $H(t) \ge -\ln|t|+b$ and $U(t,0) \ge b-\frac{1}{2}\ln|t|-H(t)$ for small values of $t$.
    \end{enumerate}
\end{theorem}

Therefore we may derive regular radiation cosmological models with a
function $U$ which grows for large values of $|t|$ and $x$ and a
function $H$ that does not grow or decrease too quickly.

\end{itemize}

\section{Conclusions}

We have derived sufficient conditions for an inhomogeneous spacetime
filled with pure radiation to be singularity-free in the sense of
causal geodesic completeness. The conditions are fairly easy to
implement in terms of a growing function $H$ and a solution to the
plane-wave equation with very few restrictions. This is much the same 
as it happened for stiff perfect fluids \cite{stiffgen}. Regular
models are pretty general among radiation spacetimes.

Pure radiation fields with positive energy density fulfill weak,
strong and dominant energy conditions.  Since these models have a
function $t$ with timelike gradient everywhere, they are causally
stable \cite{HE} and hence they satisfy every weaker causality
condition. For instance, they do not contain closed causal curves.
The only possibility to circumvent the powerful singularity theorems
is the absence of closed trapped surfaces.

It has already been shown that singularity-free models are common
among stiff perfect fluids and pure radiation inhomogeneous spacetimes.
It remains to be seen if these results can be extended to other
matter contents and symmetries.

\section*{Acknowledgements}

The present work has been supported by Direcci\'on General de
Ense\~nanza Superior Project PB98-0772. The authors wish to thank
  F. Navarro-L\'erida for valuable discussions.


\begin{thebibliography}{99.}

\bibitem{wide} L. Fern\'andez-Jambrina, L.M. Gonz\'alez-Romero, Phys.
Rev. \textbf{D66} (2002) 024027 [arxiv: gr-qc/0402119].

\bibitem{seno} J.M.M. Senovilla, Phys. Rev. Lett. \textbf{64} (1990) 
2219.\\
F.J. Chinea, L. Fern\'andez-Jambrina, J.M.M. 
Senovilla, {Phys. Rev} \textbf{D45} (1992) 481.

\bibitem{ruiz} E. Ruiz, J.M.M. Senovilla, Phys. Rev \textbf{D45} (1992) 1995.

\bibitem{Sep} M. Mars, Class. Quantum Grav. \textbf{12} (1995) 2831. 

\bibitem{dadhich} N. Dadhich: ``On the uniqueness of the singularity
free family of inhomogeneous cosmological models''.  In:
\textit{Proceedings of the Spanish Relativity Meeting on Inhomogeneous
Cosmological Models}, ed.  by A. Molina, J.M.M. Senovilla, Singapore
1995.

\bibitem{Diag} M. Mars, {\it Phys. Rev} {\bf D51} (1995) R3989.

\bibitem{Jerry} J.B. Griffiths, J. Bi\v{c}ak, Class. Quantum 
Grav. {\bf 12} (1995) L81.

\bibitem{leo}L. Fern\'andez-Jambrina, Class. Quantum Grav. \textbf{14} (1997) 3407
[arxiv: gr-qc/0404017].

\bibitem{grg} J.M.M. Senovilla, Gen. Rel. Grav. \textbf{30} (1998) 701.

\bibitem{rao} J. Krishna Rao, Proc. Nat. Inst. Sci. India A 
\textbf{30} (1964) 439.

\bibitem{HE} S.W. Hawking, G.F.R. Ellis, \textit{The large scale
structure of space-time}, Cambridge University Press, Cambridge 1973.

\bibitem{mpla} L. Fern\'andez-Jambrina, L.M. Gonz\'alez-Romero,
Mod. Phys. Lett. \textbf{A19} (2004) 583. [arxiv: gr-qc/0402124]	

\bibitem{manolo} L. Fern\'andez-Jambrina, L.M. Gonz\'alez-Romero,
Class. Quantum Grav. \textbf{16} (1999) 953 [arxiv: gr-qc/9812039].\\
L. Fern\'andez-Jambrina,  Journ. Math. Phys. 
\textbf{40} (1999) {4028} [arxiv: gr-qc/9906030].

\bibitem{stiffgen} L. Fern\'andez-Jambrina, L.M. Gonz\'alez-Romero, Journ. Math.
 Phys. \textbf{45} (2004) 2113 [arXiv: gr-qc/0405013].
\end{thebibliography}
\end{document}